\documentclass[conference]{IEEEtran}
\IEEEoverridecommandlockouts
\usepackage{cite}
\usepackage{amsmath,amssymb,amsfonts}
\usepackage{algorithm}
\usepackage{algpseudocode}
\usepackage{graphicx}
\usepackage{textcomp}
\usepackage{xcolor}
\def\BibTeX{{\rm B\kern-.05em{\sc i\kern-.025em b}\kern-.08em
    T\kern-.1667em\lower.7ex\hbox{E}\kern-.125emX}}
\begin{document}

\title{An Edge Architecture Oriented Model Predictive Control Scheme for an Autonomous UAV Mission}

\author{Achilleas Santi Seisa, Sumeet Gajanan Satpute, Björn Lindqvist, and George Nikolakopoulos%

\thanks{This work has been partially funded by the European Unions Horizon 2020 Research and Innovation Programme AERO-TRAIN under the Grant Agreement No. 953454.
}
\thanks{The Authors are with the Robotics and AI Team, Department of Computer, Electrical and Space Engineering, Lule\aa\,\,University of Technology, Lule\aa\,\,}
\thanks{Corresponding Author's email: {\tt\small achsei@ltu.se}}
}
\maketitle

\begin{abstract}
In this article the implementation of a controller and specifically of a Model Predictive Controller (MPC) on an Edge Computing device, for controlling the trajectory of an Unmanned Aerial  Vehicle (UAV) model, is examined. MPC requires more computation power in comparison to other controllers, such as PID or LQR, since it use cost functions, optimization methods and iteratively predicts the output of the system and the control commands for some determined steps in the future (prediction horizon). Thus, the computation power required depends on the prediction horizon, the complexity of the cost functions and the optimization. The more steps determined for the horizon the more efficient the controller can be, but also more computation power is required. Since sometimes robots are not capable of managing all the computing process locally, it is important to offload some of the computing process from the robot to the cloud. But then some disadvantages may occur, such as latency and safety issues. Cloud computing may offer "infinity" computation power but the whole system suffers in latency. A solution to this is the use of Edge Computing, which will reduce time delays since the Edge device is much closer to the source of data. Moreover, by using the Edge we can offload the demanding controller from the UAV and set a longer prediction horizon and try to get a more efficient controller.
\end{abstract}


\section{Introduction}
\label{Introduction}
In the future robots will have to complete more complex tasks and the requirements for an autonomous capability will increase. In that context, robots will need to have more computational power that many times and based on the mission's complexity, this will not be possible to happen locally on-board the robots' processors. These cases are the exactly ones where edge oriented control architectures are needed to provide the desired high computation power and bandwidth, while retaining an overall low latency. In this article, we propose an edge computing architecture for offloading the model predictive controller from an UAV to the Edge. By offloading the MPC to the Edge we will be able to use Edge resources and push MPC capabilities to the limits.

Edge Computing is an existing technology with tremendous upcoming possibilities. The combination of Edge Computing and 5G can revolutionize the world of robots. However, there are some limitation and challenges which researchers try to solve in order to use these technologies universally, while some examples from the state of the art in the field can be mentioned as in~\cite{b9} where a 4 layer architecture consisted of Robot, Edge, Fog and Cloud is used for offloading the localization and mapping operations. A Search Planner algorithm using Deep Learning is designed at the Edge for UAVs in~\cite{b10}. In~\cite{b11} the computational and storage resources are moved to the Edge and the Cloud for Deep Robot Learning including object recognition, grasp planning, localization etc. Finally, in~\cite{b12} a Fog architecture is proposed for communicating and controlling robots.

In all the articles mentioned above, the Edge is used mainly for offloading the demanding Artificial Intelligence, Machine or Deep Learning algorithms. On the other hand, there are not many articles presenting attempts of offloading controllers to the Edge from an automatic control systems approach. In~\cite{b7} it has been presented an application containing a remote controller that it is able to run in a Mobile Edge server in a form of Docker Container. These applications are controlling two robotic arms for a cooperation task in an industrial environment. In~\cite{b8} a switching multi-tier controller is presented. The switcher is choosing between a local controller, which operates as a safety controller and an Edge controller, which runs more sophisticated algorithms with optimal performance. In the works in~\cite{b1,b2,b3,b4,b5,b6} researchers have suggested offloading the MPC on the Edge. However, these articles focused mainly on evaluating the related latency times, the time delays and the related uncertainty for several cases. The controller in~\cite{b4} and~\cite{b3} is composed with the combination of a LQR (locally) and MPC (Edge) control, while in~\cite{b2} and~\cite{b5} two MPCs are implemented, one on a local Edge and one on a Cloud. In~\cite{b6} a variable horizon strategy for a cloud-based MPC is presented and in~\cite{b1} a remote MPC is used to control a ball and bean system.

\begin{figure}[htbp]
	\centering
	\includegraphics[width=\columnwidth]{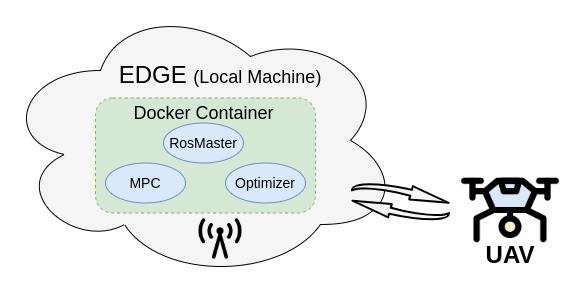}
  	\caption{System Architecture using Docker Image on the Edge}
  	\label{fig:kubernetes_architecture}
\end{figure}

The main contribution of this work is to establish an edge architecture oriented Model Predictive Control scheme for enabling a fully autonomous UAV mission. As the need of autonomous capability increases, algorithms get more and more complex and computational heavy, more hardware on board may be required and UAVs need to be lighter in order to operate longer and more efficient. Under these requirements, we were motivated to implement a platform that will meet some of these needs and will offer new capabilities. By using the Edge not only we can offload the computational demanding operations there and use lighter processors on UAVs, but also we can use the great Edge resources for computational and storage purposes. In this article we will present a novel Edge architecture to offload the MPC to the Edge and to evaluated this architecture's capabilities in terms of latency and computational power as well as overall tracking of the desired trajectories for the UAV. Towards this novel implementation, for the Edge we have used a local machine where the MPC is deployed inside a Docker container and is sending the command signals to the UAV model, the architecture of which is shown in Fig.~\ref{fig:kubernetes_architecture}

The rest of the article is structured as it follows. In Section~\ref{MPC} the Model Predictive Control scheme is presented and in Section~\ref{system_architecture} the overall system's edge architecture is established, while we also provide a detailed analysis through the parameters of the controller and we explain the need of offloading the controller to the Edge. In Section~\ref{Simulation} multiple experimental results, with a hardware in the loop approach, are presented that prove the overall efficacy of the proposed scheme, while we analyse and evaluate the usage of the Edge in matters of latency and computation capabilities. Finally, in Section~\ref{Conclusion}, the future research directions and the conclusions are derived. 
%
\section{Model Predictive Control}
\label{MPC}

\subsection{UAV Kinematics}
\label{kinematics}
Model Predictive Control has been widely used in both research and industrial environments. In this article MPC is implemented for following the desired trajectory of a UAV, where the platform is considered as a six Degree of Freedom robot with a fixed body frame and its kinematic model can be described by Eq.~\ref{eq:kinematics} in body frame as in~\cite{b13,b14}.

\begin{align}
&\dot{p}(t) = v_{z}(t) \nonumber\\
&\dot{v}(t) = R_{x,y}(\theta,\phi) \begin{bmatrix} 0\\ 0\\ T\end{bmatrix} + \begin{bmatrix} 0\\ 0\\ -g\end{bmatrix} - \begin{bmatrix} A_{x} & 0 & 0\\ 0 & A_{y} & 0\\ 0 & 0 & A_{z}\end{bmatrix}u(t) \label{eq:kinematics}\\
&\dot{\phi}(t) = \frac{1}{\tau_{\phi}} (K_{\phi} \phi_{d}(t) - \phi(t)) \nonumber\\
&\dot{\theta}(t) = \frac{1}{\tau_{\theta}} (K_{\theta} \theta_{d}(t) - \theta(t)) \nonumber
\end{align} 

In Eq.~\ref{eq:kinematics} $p = [p_{x}, p_{y}, p_{z}]^{T}$ is the position and $v = [v_{x}, v_{y}, v_{z}]^{T}$ is the linear velocity referenced in the global frame. $R(\phi(t), \theta(t)) \in SO(3)$ is the rotation matrix that describes the attitude in Euler form. $\phi$ and $\theta \in [-\pi, \pi]$ are the roll and pitch angles along the $x^{\mathbb{W}}$ and $y^{\mathbb{W}}$ axes respectively, while $\phi_{d}$ and $\theta_{d} \in R$ and $T \geq 0$ are the desired inputs values to the system in roll, pitch and the total thrust. The above model assumes that the acceleration is only dependant on the magnitude and angle of the thrust vector, produced by the motors, as well as the linear damping terms $A_{x}, A_{y}, A_{z} \in R$ and the gravity of earth $g$. The attitude terms are modeled as a first-order system between the attitude (roll/pitch) and the desired $\phi_{d}$ and $\theta_{d} \in R$, with gains $K_{\phi}$ and $K_{\theta} \in R$ and time constants $\tau_{\phi}$ and $\tau_{\theta} \in R$. It is also assumed that the UAV is equipped with a lower-level attitude controller that takes thrust, roll and pitch commands and provides motor commands for the UAV.
\subsection{Cost Function}
\label{cost_function}
For the cost function, the UAV's state vector is represented as $x = [p, v, \phi, \theta]^{T}$ and the control input as $u = [T, \phi_{d}, \theta_{d}]^{T}$. The system has a sampling time of $\delta_{t} \in \mathbb{Z}^{+}$ using forward Euler method for each time instance $(k+1|k)$, while this discrete model is used as the prediction model of the MPC. The prediction considers the specified number of steps into the future, which is called prediction horizon and it is represented as $N$. An related optimizer is tasked with finding an optimal set of control actions, defined by the cost minimum of this cost function, by associating a cost to a configuration of states and inputs at the current time and in the prediction. The predicted states at the time step $k+j$, produced at the time step $k$ are represented as $x_{k+j|k}$.
The corresponding control actions are represented as $u_{k+j|k}$. Also $x_{k}$ and $u_{k}$ represent the full predicted states and corresponding control inputs along the prediction horizon correspondingly. The objective of the controller is to navigate to the desired position and deliver smooth control inputs. The cost function is presented in Eq.~\ref{eq:cost_funtion}.

\begin{align}
&J = \sum_{j=1}^{N} \underbrace{(x_{d} - x_{k+j|k})^{T} Q_{x} (x_{d} - x_{k+j|k})}_{state \quad cost} \nonumber\\
&+ \underbrace{(u_{d} - u_{k+j|k})^{T} Q_{u} (u_{d} - u_{k+j|k})}_{input \quad cost} \label{eq:cost_funtion}\\
&+ \underbrace{u_{k+j|k} - u_{k+j-1|k})^{T} Q_{\delta u} (u_{k+j|k} - u_{k+j-1|k})}_{control \quad actions \quad smoothness \quad cost} \nonumber
\end{align} 

\begin{figure*}[ht!]
	\centering
	\includegraphics[width=0.95\textwidth]{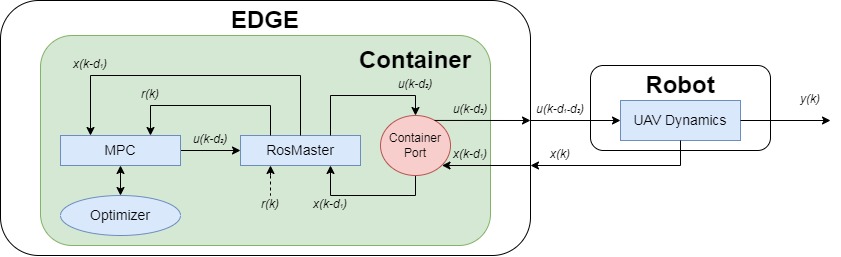}
  	\caption{Block Diagram of the Edge Architecture for the UAV-MPC System}
  	\label{fig:block_diagram}
\end{figure*}
where $Q_{x} \in \mathbb{R}^{8x8}$ is the matrix for the state weights, $Q_{u}$ is the matrix for the input weights and $Q_{\delta u} \in \mathbb{R}^{3x3}$ is the matrix for the input rate weights. In Eq.~\ref{eq:cost_funtion} the first term describes the state cost, which is the cost associated with deviating from a certain desired state $x_{d}$. The second term describes the input cost that penalizes a deviation from the steady-state input $u_{d} = [g, 0, 0]$ and represent the inputs that describe hovering. The final term is added to guarantee that the control actions are smooth. That is achieved by comparing the input at $(k+j-1|k)$ with the input at $(k+j|k)$ and penalizing the changing of the input from one time step to the next one, with $N \in N^{+}$ to denote the control Horizon of the MPC. In this case, we evaluate the overall behavior of the MPC scheme by changing the values of the horizon and the execution rate and measuring the overall latency times from the proposed edge architecture. while the motivation of experimenting with these parameters is described in Sections~\ref{MPC_Horizon} and ~\ref{MPC_Rate}.

\subsubsection{MPC Horizon}
\label{MPC_Horizon}
MPC is optimizing a finite prediction horizon but is using only the next time-slot, while this process is executed again, repeatedly at every step. By utilizing the MPC, we optimize the current step, while keeping the future time-slots in account. By increasing the finite number of the prediction horizons the process becomes more computational heavy, thus more resources are required. On the other hand, since MPC can predict the change of dependent variables, an increased prediction horizon can make better predictions and predict changes faster. The limited computational capability of UAVs' on-board processors are setting some boundaries on how much we can increase the number of steps for the prediction horizon. To overcome this constrains, we use the computational power of the Edge, where we are able to increase the horizon and evaluate the results in Section~\ref{Simulation}.

\subsubsection{MPC Rate}
\label{MPC_Rate}
Another parameter's value that we were able to increase on the Edge was the value of the MPC Rate. That is how fast the MPC is executed. A faster MPC means a faster control system that can generate commands, to be send to the UAV rotors, faster. This would be handful in situations where we want the system to respond rapidly. This characteristic is of great importance especially in the case of aerial vehicle systems. As in the previous cases, the computational capabilities of the UAVs' processors set some limits that we were able to overcome since we used the Edge for these processes. The processors of some UAVs that we use have the capability to run the MPC at 20Hz, but by using the Edge we were able to increase the rate up to 100Hz. The results of increasing the MPC Rate are shown in Section~\ref{Simulation}.
\section{Edge Architecture}\label{system_architecture}
The Container is a unit of software that runs code and all the dependencies so the application deployed on the container will run quickly and reliably from any computer. In our case, the application Docker container runs the MPC on the Edge, while the UAV model is implemented on a local computer. The local machine used as an Edge server and the local computer running the UAV model must share the same network. In the propose architecture we have utilized an Ubuntu 20.04 Container Image running ROS Noetic for deploying the MPC on the Edge, and the same operating system and software on our local computer.

In this architecture, MPC commands are sent from the Edge to a local computer, which runs a simulation of a UAV, as shown in Fig.~\ref{fig:kubernetes_architecture} that represents the basic structure of the system. Furthermore, in this implementation, we have utilized the TCP protocol for the communication between the Edge and the local computer, which are under the same network, while we ran RosMaster, MPC and Optimizer ROS Nodes on the Edge. The simulation node ran on Gazebo environment on our local computer. The architecture and the block diagram of the system are shown in Fig.~\ref{fig:block_diagram}, while the ROS operation is shown in Fig.~\ref{fig:ros_architecture}. 

The parameters of the system are depicted in Fig.~\ref{fig:block_diagram}, where $r(k)$ is the reference input signal for the desired trajectory and $x(k)$ is the states signal generated by the UAV dynamics. In the proposed architecture, we feedback the states signal to the MPC. Since there is latency from the time instance that the UAV dynamics, on the local computer, generate the states signal to the time instant that the signal gets to the MPC, on the Edge (round trip delay), the states signal arriving to the MPC is delayed and denoted as $x(k-d_{1})$, where $d_{1}$ represents the time delay. The reference and the states signal describe the desired pose and the real pose of the UAV respectively. $u(k-d_{2})$ is the command signal generated by the MPC. The value $d_{2}$ represents the MPC execution time that is depending on the MPC rate and the computational process. Again, since the command signal has to travel from the Edge to the local computer, where the UAV model is implemented, the command signal arriving to the UAV can be denoted as is $u(k-d_{1}-d_{2})$, while the command signal is the necessary thrust for each one of the rotors for the desired trajectory. Finally, $y(k)$ is the output of the system which denotes the $x,~y,~z,~yaw$ values of the real pose of the UAV.
\section{Simulation Results and Evaluation} \label{Simulation}
\begin{figure}[ht!]
	\centering
	\includegraphics[width=\columnwidth]{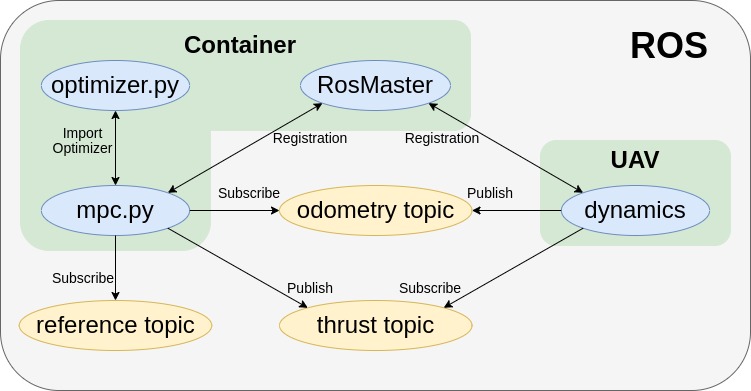}
  	\caption{ROS Architecture showing: a)ROS nodes in blue, and b)ROS topics in yellow. The lines point out the relationship between topics and nodes}
  	\label{fig:ros_architecture}
\end{figure}

In this Section we evaluate the system and present the simulation results with a hardware in the loop approach (edge architecture and software model for the aerial platform. In more details, we used an Edge Server located in Luleå, Sweden. The Edge Server is providing the needed computational resources with low latency. In Fig.~\ref{fig:ros_architecture} the ROS structure is depicted, where we import the Optimizer into the MPC Node, which subscribes to the odometry topic (measured position depended on UAV real position) and reference topic (reference position depended on desired trajectory) and publishes the commands to the thrust topic (MPC output value of thrust for desired trajectory), while the UAV simulation subscribes to the thrust topic to receive the commands and publishes its position to the odometry topic. In order for the MPC and the UAV dynamics Nodes to exchange data by publishing and subscribing to topics, they have to register to the RosMaster, which is running on the Edge.

The sequence of the operation is described in Alg.~\ref{alg:cap}, where the MPC controller is taking care of the UAV to follow the desired trajectory with an Error Tolerance of $0,4m$.

\begin{algorithm}
\caption{Algorithmic Sequence of Operations }\label{alg:cap}
1. Start RosMaster Node \\
2. Run Optimizer \\
3. Start UAV Dynamics Node \\
4. UAV Reads Real Position $[x, y, z]^{T}$ from Sensors \\
5. Start UAV MPC Node \\
6. UAV Start Take Off \\
7. UAV Hovering at $[x, y, z]^{T} = [0, 0, 1]^{T}$ \\
8. Load Desired Trajectory \\
9. Calculated Way Points ($x_{ref}, \quad y_{ref}, \quad z_{ref}$) \\
10. UAV Follow the Way Points
\begin{algorithmic}
    \State $k = 0$
    \While{$x_{ref}(k) - x, \quad y_{ref}(k) - y, \quad z_{ref}(k) - z \geq 0.4m$}
        \State $goto \quad x_{ref}(k), \quad y_{ref}(k), \quad z_{ref}(k)$
        \EndWhile
    \State $k = k + 1$
\end{algorithmic}    
\end{algorithm}

For the simulation of the UAV model we used the Gazebo simulation software, while we were able to visualize the behavior of the system for each one of the different MPC horizon and rate values. Furthermore, we recorded the data from the odometry, reference and thrust topics and we evaluated the system by using the MATLAB environment of MathWorks to extract useful data and plots.

\begin{figure}[ht!]
	\centering
	\includegraphics[width=\columnwidth]{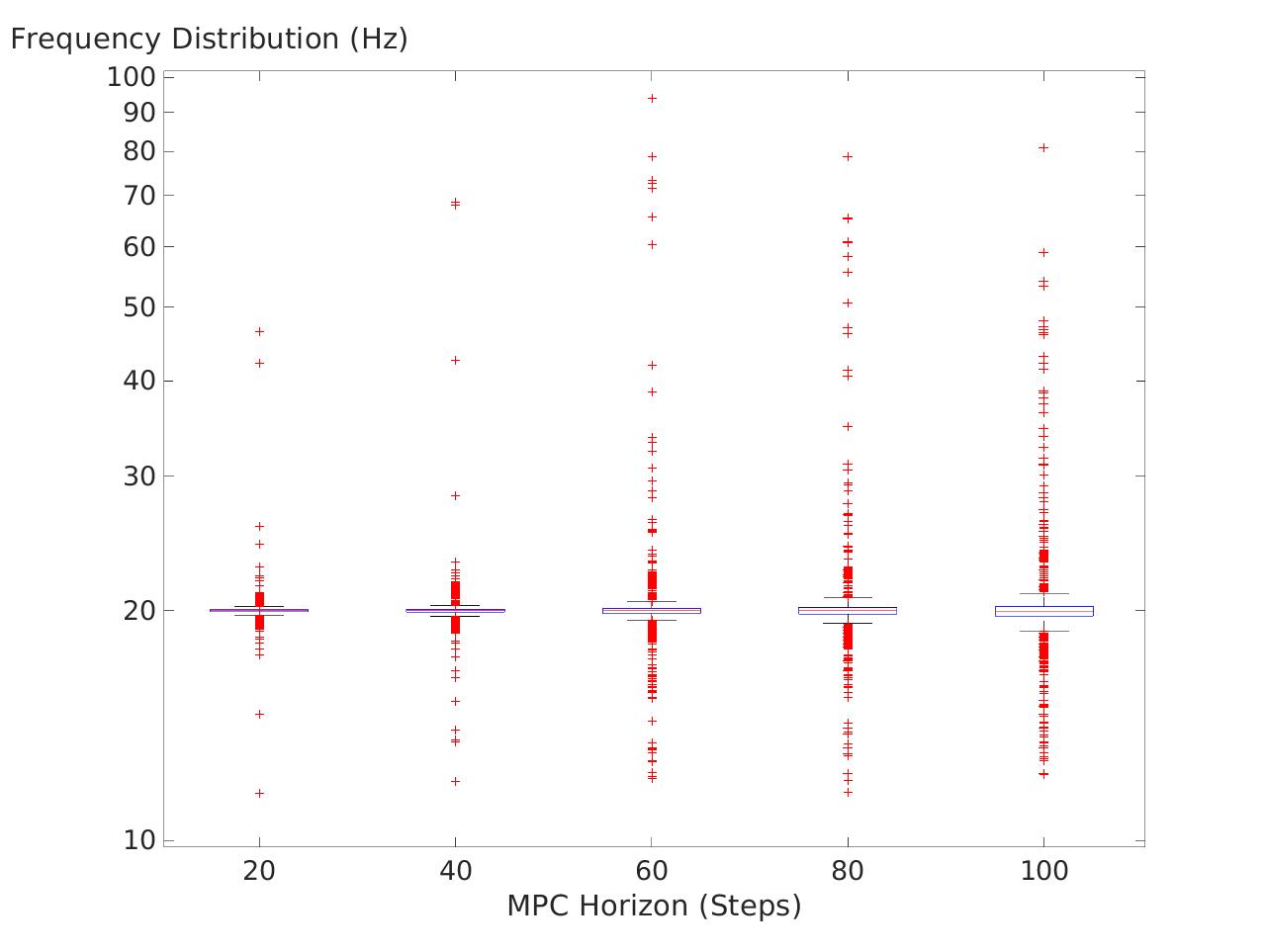}
  	\caption{Frequency distribution for different MPC Horizon values}
  	\label{fig:different_horizons}
\end{figure}

\begin{figure}[ht!]
	\centering
	\includegraphics[width=\columnwidth]{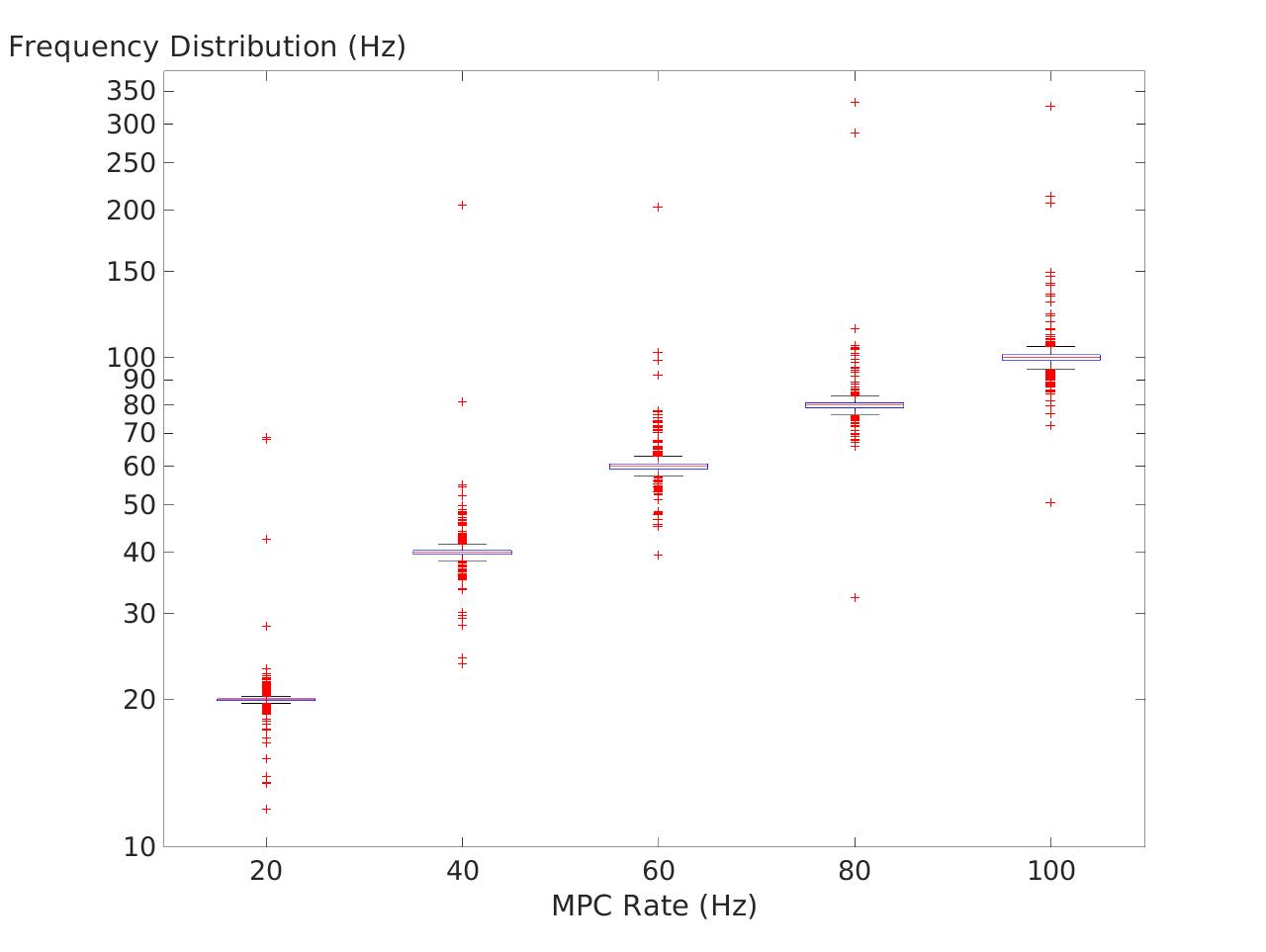}
  	\caption{Frequency distribution for different MPC Rate values}
  	\label{fig:different_mpc_rate}
\end{figure}

To evaluate our proposed architecture, we completed a series of simulation tests. Our goal was to evaluate the system in terms of latency and computational capacity. To achieve that we choose two different desired trajectories, a circular and a spiral. We determined different MPC horizons and rates and we run several tests. For each test we measured the time delays and we evaluated the responses. 
The objective were to increase the MPC horizon and rate in values that the local processor would not be able to handle. In Fig.~\ref{fig:different_horizons} and Fig.~\ref{fig:different_mpc_rate} we depict the latencies for different values of horizons and rates respectively. The Table~\ref{table_dif_hor} and Table~\ref{table_dif_rate} present some information regarding the latencies.

\begin{table}[htbp]
	\centering
	\caption{Latency in milliseconds for different horizon steps and fixed rate at 40 Hz}
	\begin{tabular}{ |c||c|c|c|c|c| } 
 	\hline
 	\centering
 	\shortstack{\\$Horizon Steps$} & $20$ & $40$ & $60$ & $80$ & $100$\\
 	\hline
 	\centering
 	\shortstack{\\$Minimum$} & 21.518 & 14.583 & 10.647 & 12.691 & 12.362\\
 	\hline
 	\centering
 	\shortstack{\\$Lower Adjacent$} & 49.348 & 49.158 & 48.582 & 48.054 & 47.217\\
 	\hline
 	\centering
 	\shortstack{\\$25^{th} Percentile$} & 49.843 & 49.785 & 49.642 & 49.511 & 49.363\\
 	\hline
 	\centering
 	\shortstack{\\$Median$} & 50.002 & 49.998 & 50.008 & 50.010 & 50.033\\
 	\hline
 	\centering
 	\shortstack{\\$75^{th} Percentile$} & 50.175 & 50.208 & 50.349 & 50.492 & 50.796\\
 	\hline
 	\centering
 	\shortstack{\\$Upper Adjacent$} & 50.658 & 50.840 & 51.397 & 51.946 & 52.927\\
 	\hline
 	\centering
 	\shortstack{\\$Maximum$} & 86.705 & 83.665 & 82.910 & 86.629 & 82.007\\
 	\hline
	\end{tabular}
	\label{table_dif_hor}
\end{table}

As we can observe from Fig.~\ref{fig:different_horizons} and Table~\ref{table_dif_hor}, by increasing the value of the MPC horizon, the median time delay remains almost the same, but we have more deviation. This was expected, since longer horizon means more computations so the execution of the MPC might get slower in some cases. The difference of deviation between the shorter and longer chosen horizon is in single digit millisecond. 

\begin{table}[htbp]
	\centering
	\caption{Latency in milliseconds for different rates and fixed horizon at 20 steps}
	\begin{tabular}{ |c||c|c|c|c|c| } 
 	\hline
 	\centering
 	\shortstack{\\$Rate (Hz)$} & $20$ & $40$ & $60$ & $80$ & $100$\\
 	\hline
 	\centering
 	\shortstack{\\$Minimum$} & 14.583 & 4.878 & 4.931 & 3.003 & 3.062\\
 	\hline
 	\centering
 	\shortstack{\\$Lower Adjacent$} & 49.158 & 24.071 & 15.847 & 11.933 & 9.463\\
 	\hline
 	\centering
 	\shortstack{\\$25^{th} Percentile$} & 49.785 & 24.768 & 16.465 & 12.359 & 9.869\\
 	\hline
 	\centering
 	\shortstack{\\$Median$} & 49.998 & 25.006 & 16.682 & 12.503 & 10.000\\
 	\hline
 	\centering
 	\shortstack{\\$75^{th} Percentile$} & 50.208 & 25.240 & 16.868 & 12.650 & 10.141\\
 	\hline
 	\centering
 	\shortstack{\\$Upper Adjacent$} & 50.840 & 25.939 & 17.464 & 13.080 & 10.545\\
 	\hline
 	\centering
 	\shortstack{\\$Maximum$} & 83.665 & 42.261 & 25.299 & 30.854 & 19.809\\
 	\hline
	\end{tabular}
	\label{table_dif_rate}
\end{table}

In Fig.~\ref{fig:different_mpc_rate} and Table~\ref{table_dif_rate} we can see that by increasing the MPC Rate, the controller gets much faster and we do not suffer from standard deviations. This means that we can take advantage of the Edge resources and we can implement a much faster controller. As we already mentioned, this is essential for the UAV because with a fast controller, the system can respond faster. The importance of a fast controller can be shown in the case of collision avoidance, where the UAV have to respond fast in order to prevent the collision. The combination of a fast controller with a predictive behavior can help the system avoid the collision in an optimal way.

\begin{figure}[ht!]
	\centering
	\includegraphics[width=\columnwidth]{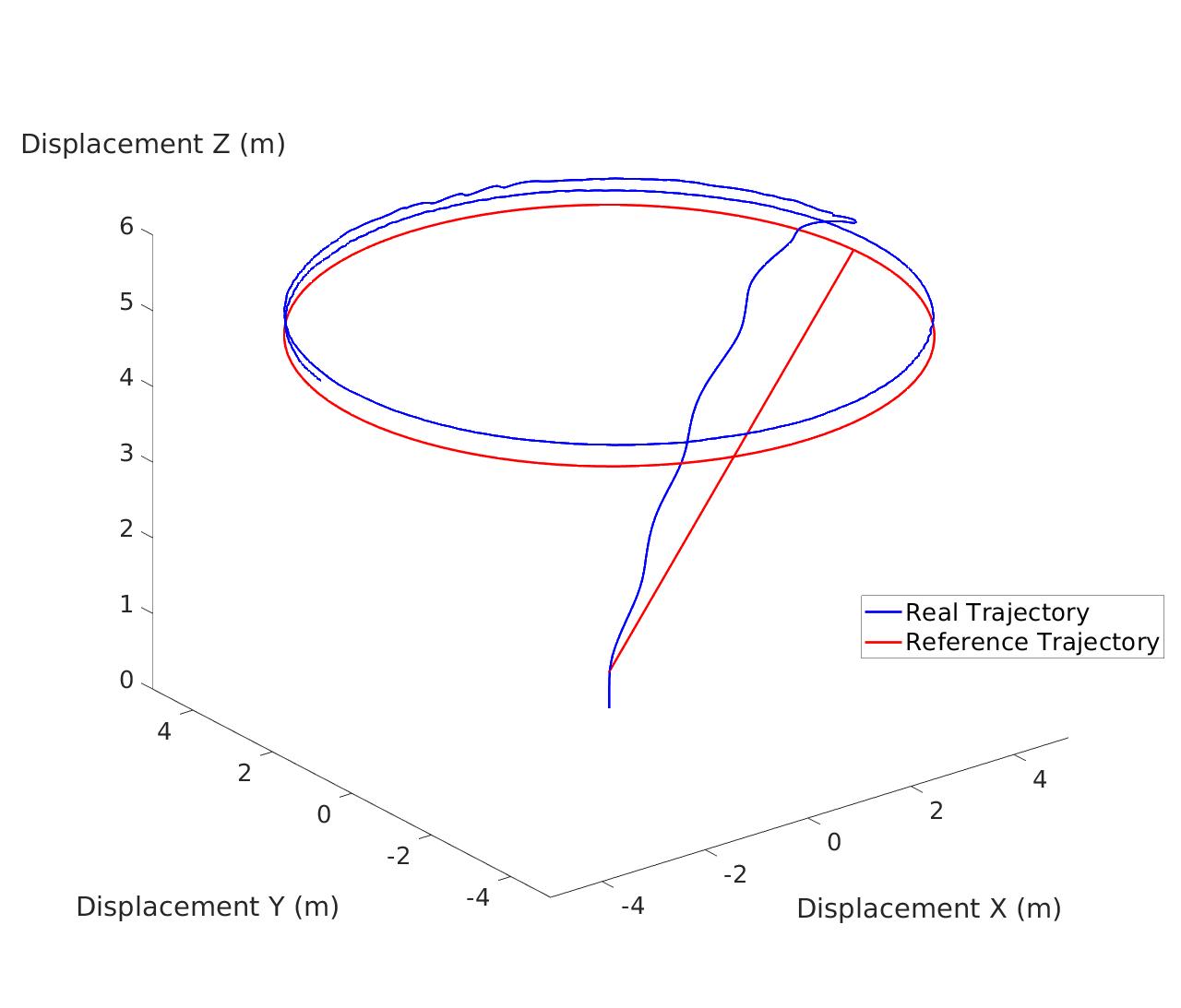}
  	\caption{3D view of the circular trajectory. Red line: reference signal. Blue line: real trajectory}
  	\label{fig:circular_trajectory_3D}
\end{figure}

\begin{figure}[ht!]
	\centering
	\includegraphics[width=\columnwidth]{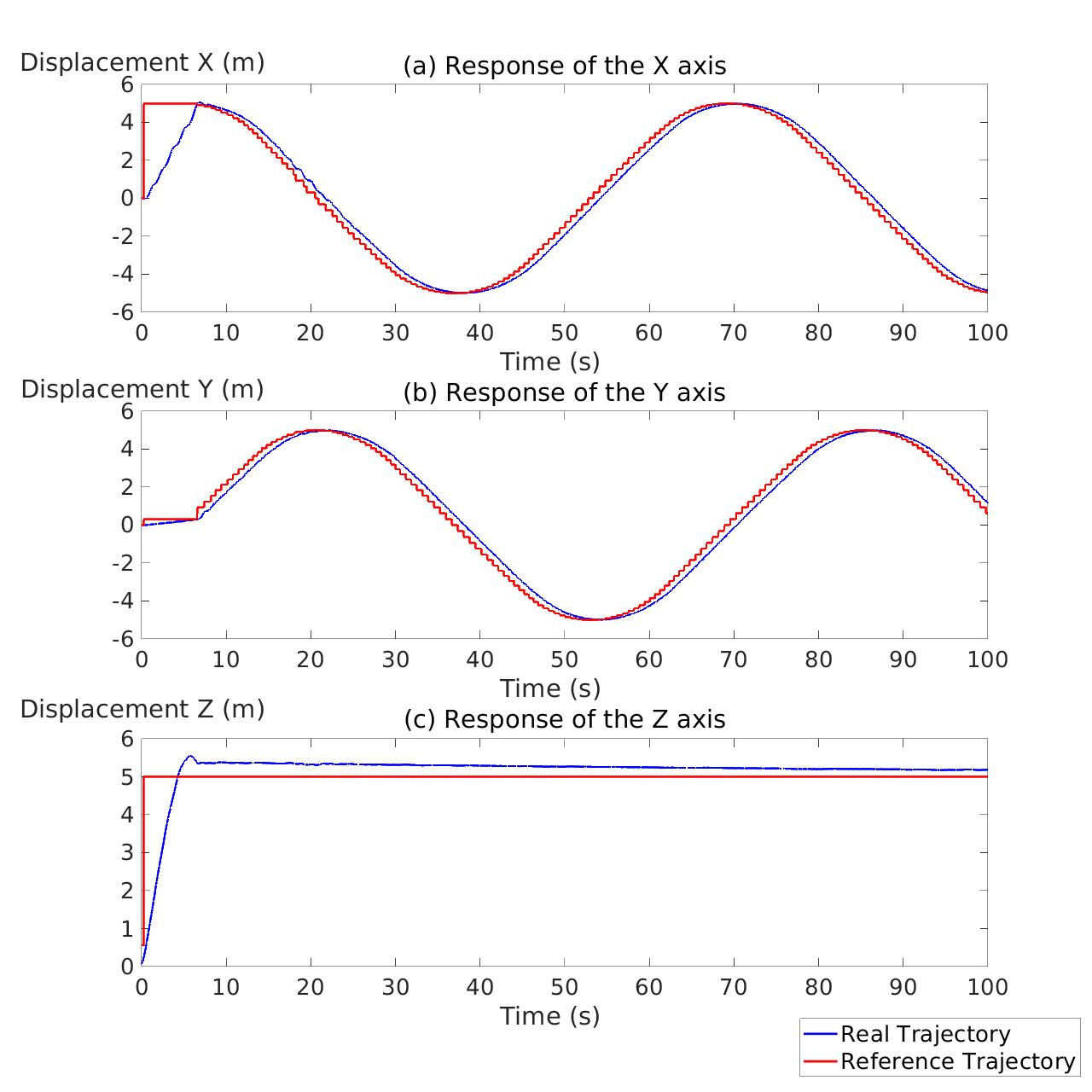}
  	\caption{Responses of the circular trajectory for each axis. a)x axis response, b)y axis response and c)z axis response}
  	\label{fig:circular_trajectory}
\end{figure}

\begin{figure}[ht!]
	\centering
	\includegraphics[width=\columnwidth]{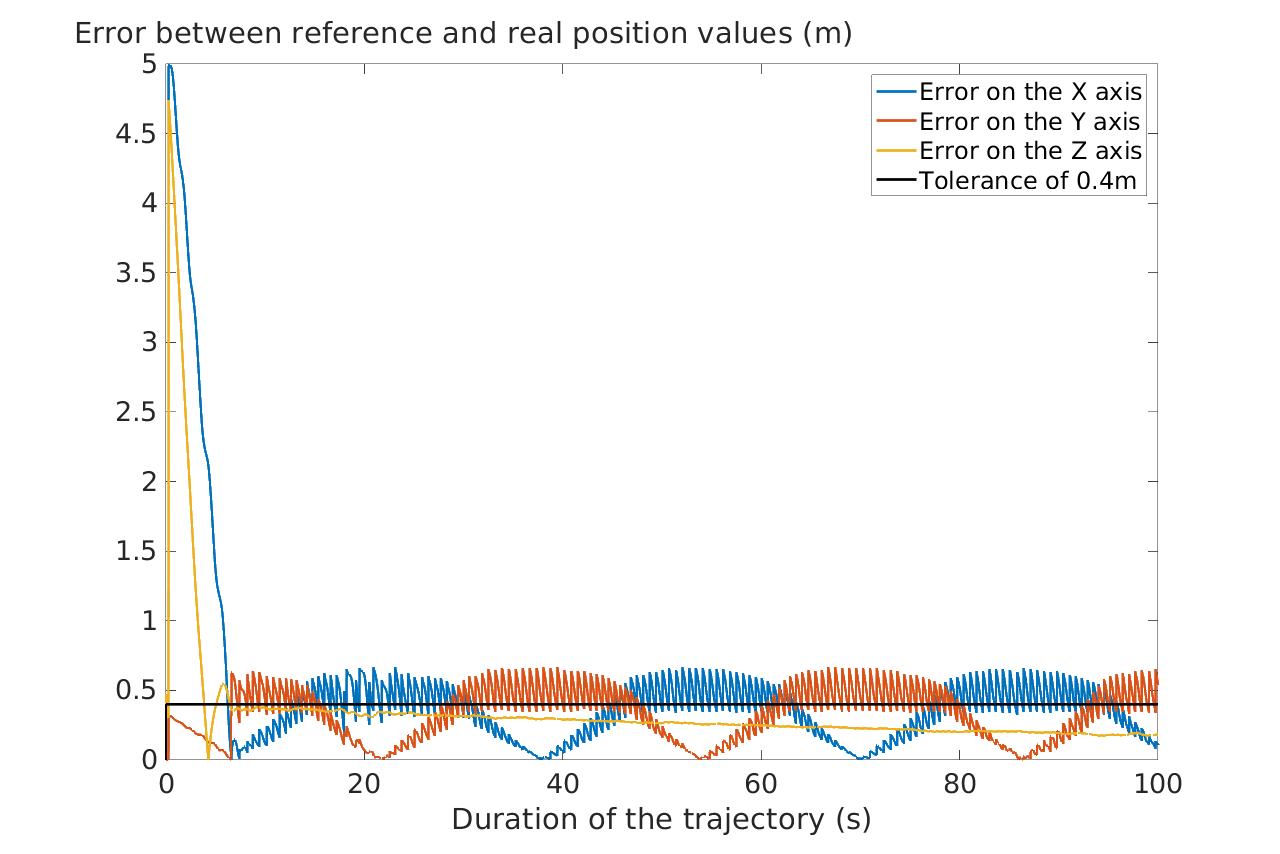}
  	\caption{Error between reference and real position on X, Y and Z axis during circular trajectory}
  	\label{fig:circular_error}
\end{figure}

\begin{figure}[ht!]
	\centering
	\includegraphics[width=\columnwidth]{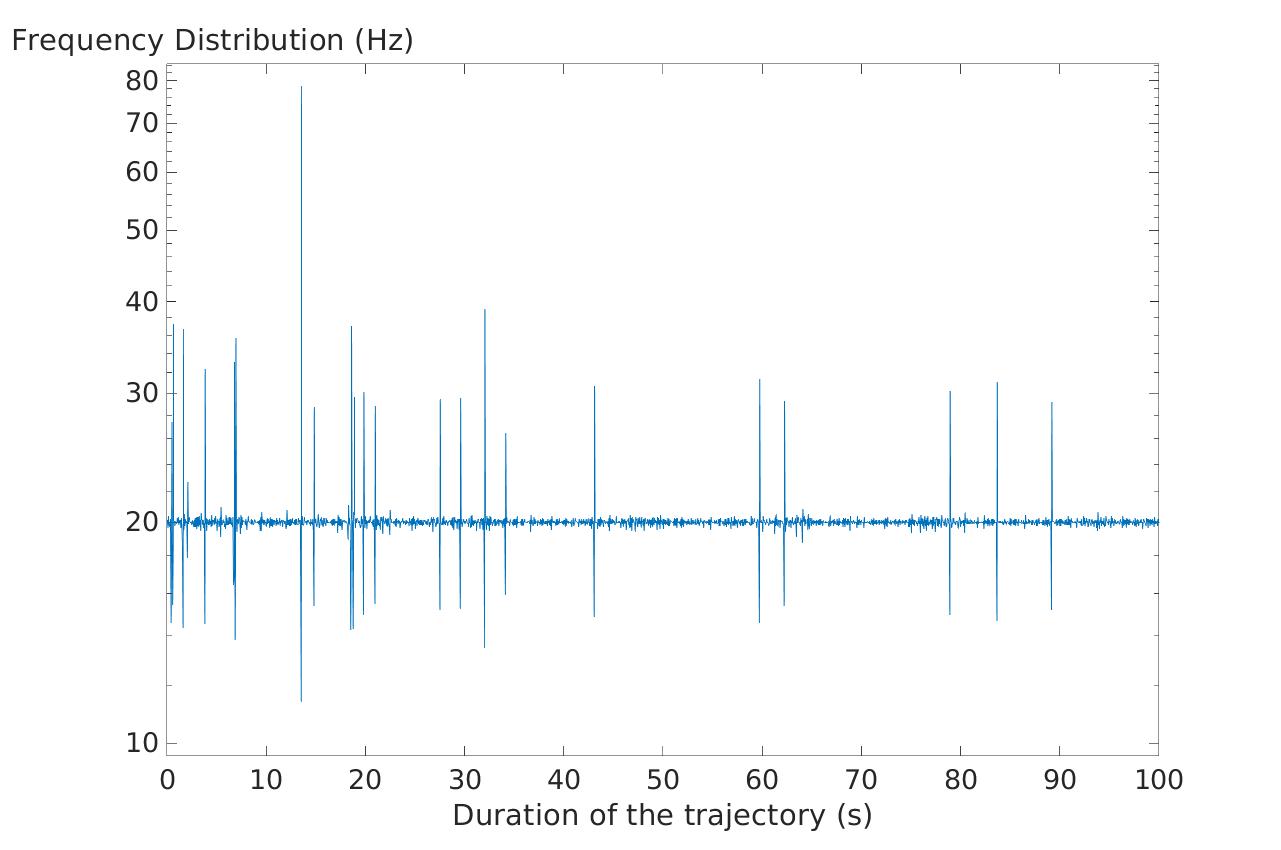}
  	\caption{Latency for each signal step of the circular trajectory}
  	\label{fig:circular_latencies}
\end{figure}

\begin{figure}[ht!]
	\centering
	\includegraphics[width=\columnwidth]{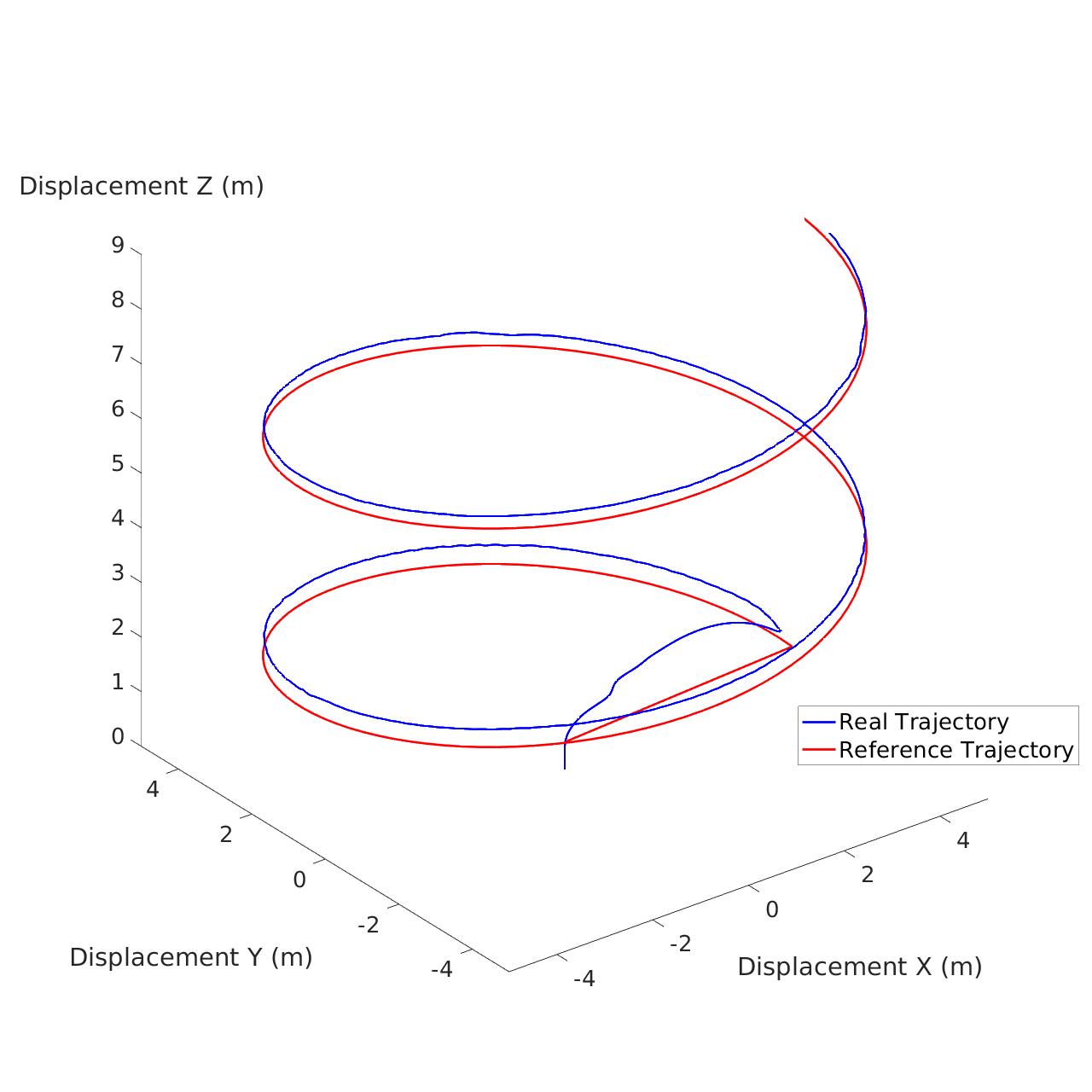}
  	\caption{3D view of the spiral trajectory. Red line: reference signal. Blue line: real trajectory}
  	\label{fig:spiral_trajectory_3D}
\end{figure}

\begin{figure}[ht!]
	\centering
	\includegraphics[width=\columnwidth]{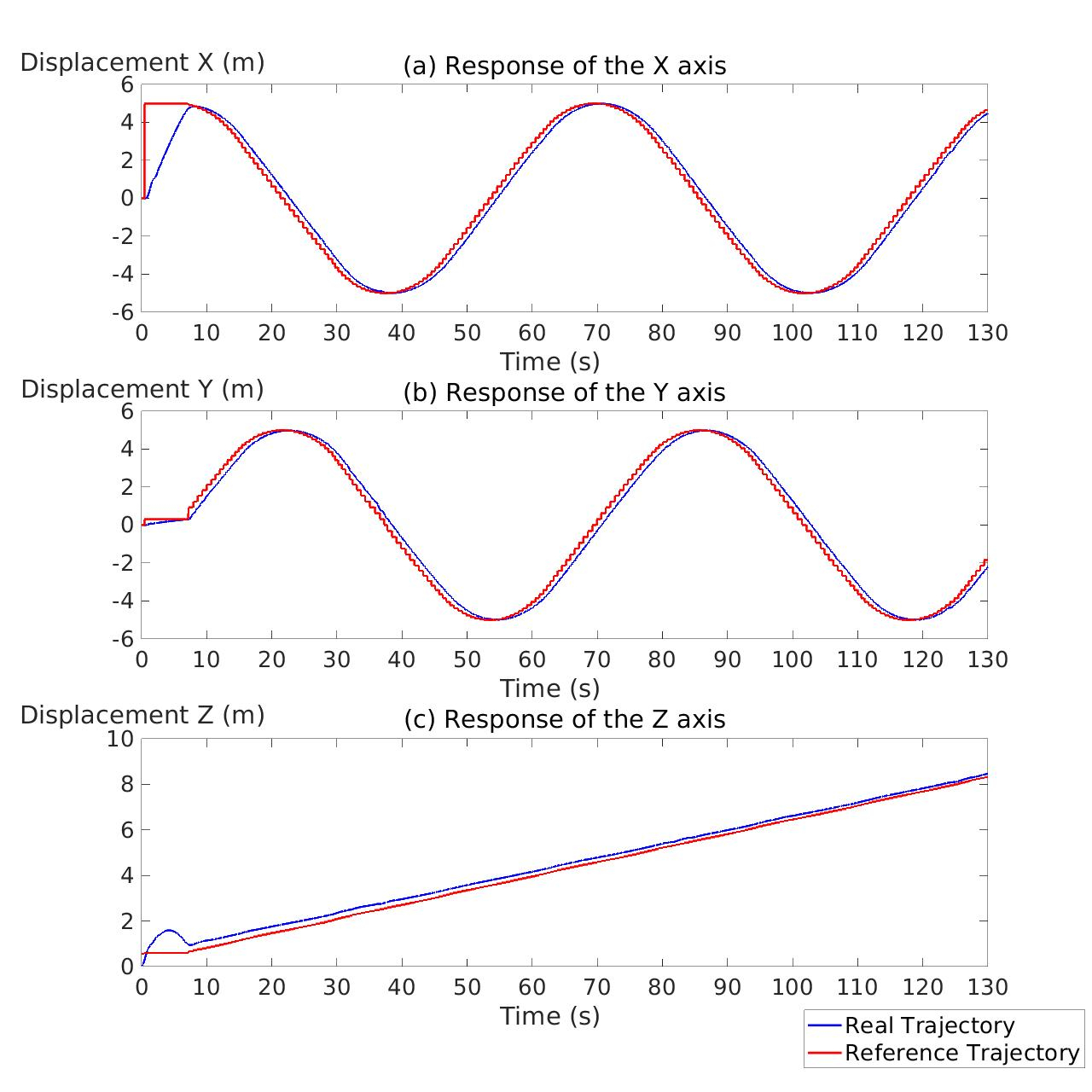}
  	\caption{Responses of the spiral trajectory for each axis. a)x axis response, b)y axis response and c)z axis response}
  	\label{fig:spiral_trajectory}
\end{figure}

\begin{figure}[ht!]
	\centering
	\includegraphics[width=\columnwidth]{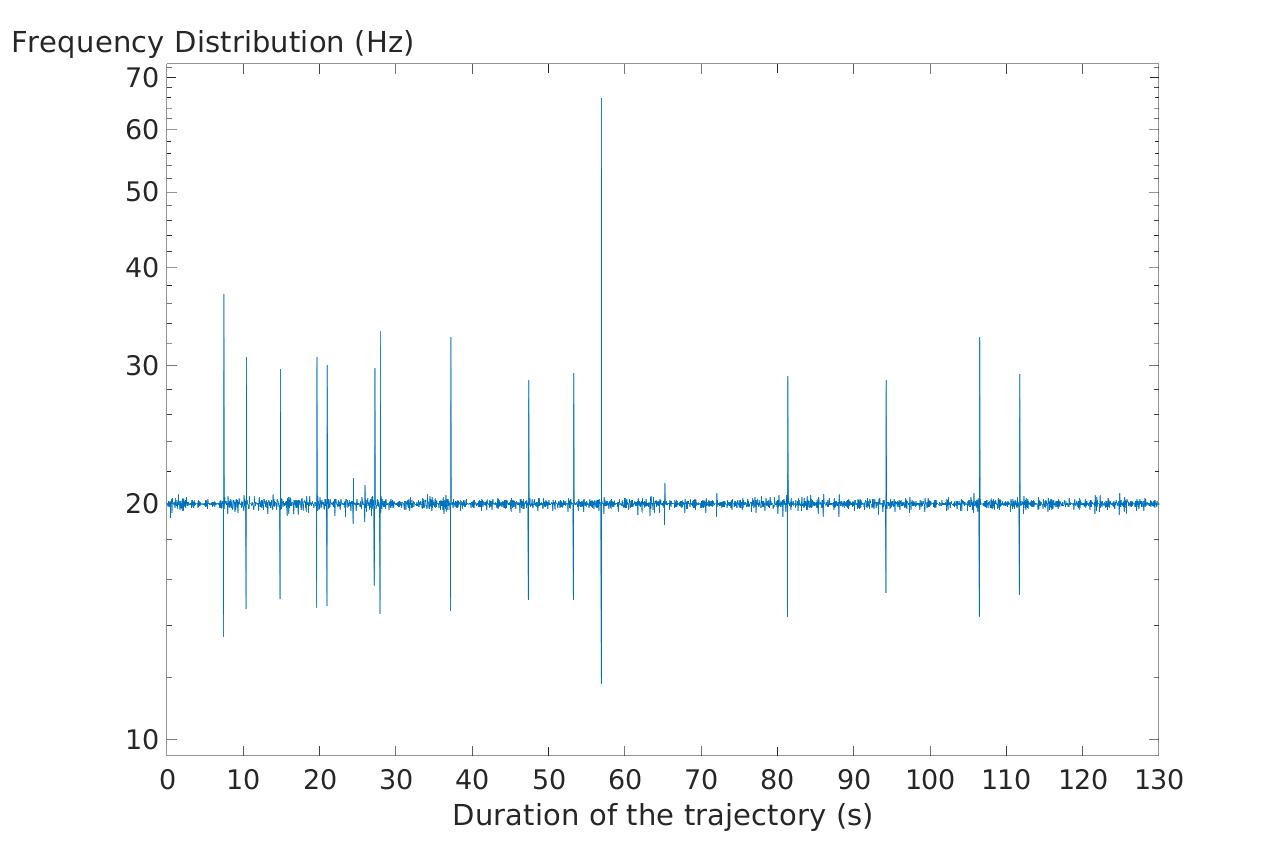}
  	\caption{Latency for each signal step of the spiral trajectory}
  	\label{fig:spiral_latencies}
\end{figure}

The round-trip for the MPC execution for each step is shown in Fig.~\ref{fig:circular_latencies}. These measurements are from the circular trajectory shown in Fig.~\ref{fig:circular_trajectory} where the duration of the circular trajectory was 100 seconds and for the spiral trajectory was 130 seconds. The MPC rate is set at 20Hz and the MPC Horizon steps are set to 20. The mean execution time is 49.95 milliseconds, which is almost the same to the MPC Rate 20Hz, which is equal to 50 milliseconds. In Fig.~\ref{fig:circular_trajectory_3D},~\ref{fig:circular_trajectory},~\ref{fig:spiral_trajectory_3D} and~\ref{fig:spiral_trajectory} the responses of the circular and spiral trajectories are depicted, where the red line represents the reference signal and the blue line the real trajectory, while in Fig.~\ref{fig:circular_latencies} and Fig.~\ref{fig:spiral_latencies} we can observe the latencies for each time step of the two different trajectories, respectively.

In Fig.~\ref{fig:circular_trajectory_3D} we are presenting the 3D responses of the UAV model when receiving the commands, for following a circular trajectory, from the Edge MPC. The blue line represents the desired trajectory and the red line represents the real trajectory of the UAV model. The real trajectory is following the desired one under the requirements with an error $\in [0, 0.4m]$. The same principals are applied for the spiral trajectory in Fig.~\ref{fig:spiral_trajectory_3D}.

Moreover, we are presenting the responses in each axe in Fig.~\ref{fig:circular_trajectory_3D} and~\ref{fig:circular_trajectory} for the circular and the spiral trajectory respectively. In these figures it is easy to observe the steady state error, which is reasonable since we applied an error tolerance of $0.4m$.

\begin{figure}[ht!]
	\centering
	\includegraphics[width=\columnwidth]{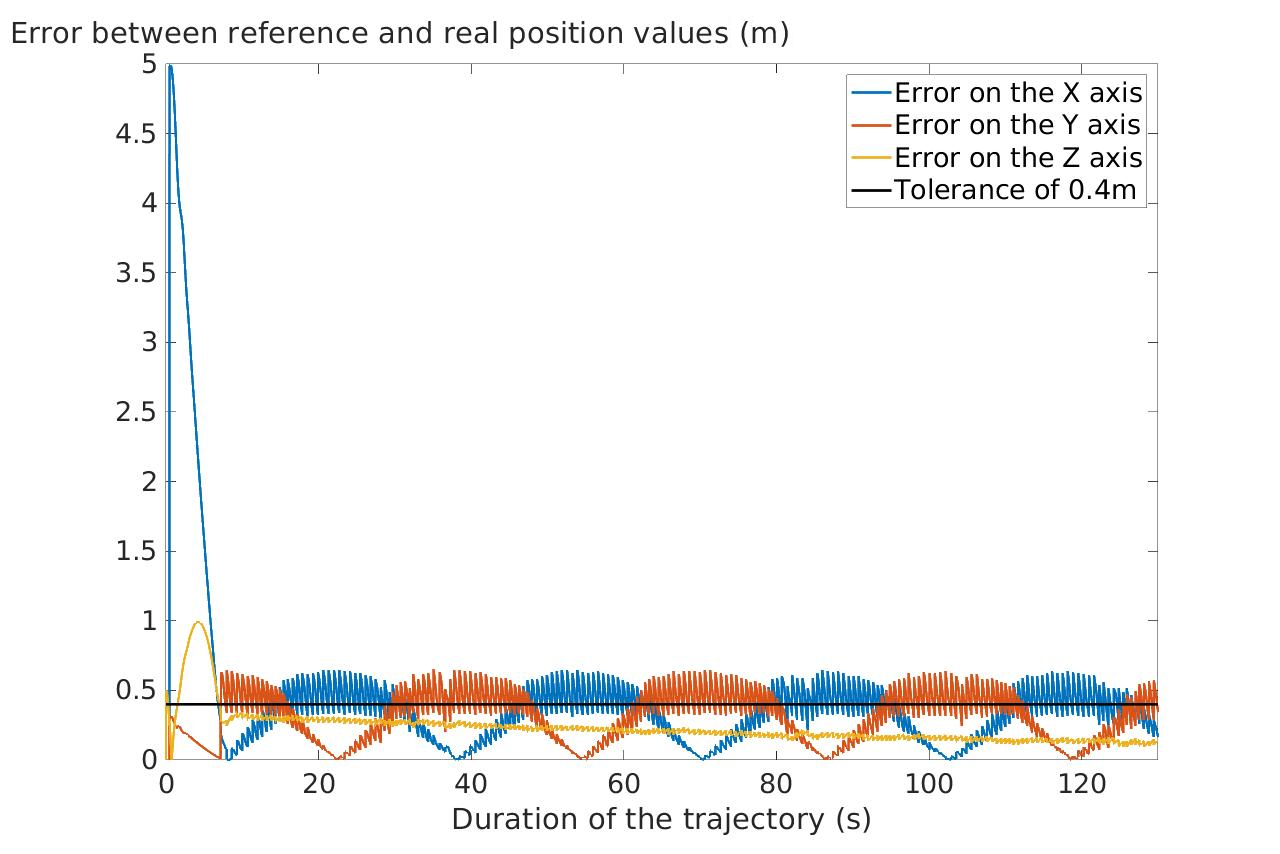}
  	\caption{Error between reference and real position on X, Y and Z axis during spiral trajectory}
  	\label{fig:spiral_error}
\end{figure}

Furthermore, the errors between the reference and real position on X, Y and Z axis during circular and spiral trajectories are depicted in Fig~\ref{fig:circular_error} and~\ref{fig:spiral_error} respectively. The controller tends to keep the error under the tolerance of 0.4m.

Finally, in Fig.~\ref{fig:circular_latencies} and Fig.~\ref{fig:spiral_latencies} we are presenting the overall latency of each time step. We calculated the latency by extracting the frequency distribution between two sequential commands. The latency as we explained in Section~\ref{system_architecture}, depends on two parameters, the MPC execution time, based on the MPC rate and computational needs, and the round trip delay. So the latency in Fig.~\ref{fig:circular_latencies} and~\ref{fig:spiral_latencies} is the sum of the latencies for each time step as shown in Eq.~\ref{eq:latencies}.

\begin{equation}
L_{total} = L_{rtd} + L_{exec}
\label{eq:latencies}
\end{equation}

where $L_{total}$ is the total latency, $L_{rtd}$ is the round-trip latency (RTL) and $L_{exec}$ is the MPC execution time latency.

\section{Conclusion} \label{Conclusion}
We were able to successfully control the trajectory of an UAV with a MPC operating on the Edge, but still there are some improvements that can be applied. Our next step would be to deploy our Docker Containers as PODs on a Cluster and use Kubernetes for orchestration. POD is the group of containers sharing storage and network resources and Kubernetes is an orchestration system for automatic application deployment, scaling and management~\cite{b15}, while this would be helpful in terms of deploying and managing our containers and it will be also possible to assign the desired resources and create services for better and more efficient management.

Over and above that, there are some further directions for potential improvements. The major issues with our system's architecture is the latency and the lack of a backup controller running locally on UAV's on-board processor. We will be able to challenge these throwbacks by using 5G network capabilities for reducing the latency by reducing the signals' travel time. Additionally, 5G will provide us more bandwidth for huge amount of data streaming. Moreover, we can have a backup computation light controller locally on the UAV that will operate only in case of network connection lost. 

Some interesting future direction will be to operate in challenging environments and scenarios where the long MPC horizon and the high MPC rate will come in use. These scenarios can be an obstacle or collision avoidance missions and in environments where the UAVs navigation is rather difficult. Additionally, we can use the capabilities of Edge for UAVs' operations for more complex tasks. Offloading these complex tasks on the Edge and having a multi agent systems cooperation in real time will be a fascinating future direction. With the development of the relevant technologies, the capabilities on the field are expected to expand tremendously.

\end{document}